\newcommand\DIMPYd{(C$_7$D$_{10}$N)$_2$CuBr$_4$}
\newcommand\DIMPY{(C$_7$H$_{10}$N)$_2$CuBr$_4$}
\newcommand\Hpip{(C$_5$H$_{12}$N)$_2$CuBr$_4$}
\newcommand\IPA{(CH$_3$)$_2$CHNH$_3$CuCl$_3$}

\documentclass[prb,twocolumn,floatfix,preprintnumbers,amsmath,amssymb]{revtex4}

\usepackage{graphicx}		
\usepackage{dcolumn}		
\usepackage{bm}					
\usepackage{enumerate} 	
\usepackage{color}
\usepackage{ulem}

\begin{document}

\title{Symmetric and asymmetric excitations of a strong-leg quantum spin ladder}

 \author{D. Schmidiger}
 \affiliation{Neutron Scattering and Magnetism,
 							Laboratory for Solid State Physics, ETH Z\"urich,
							Switzerland}

\author{P. Bouillot}
\affiliation{Department of Medical Imaging and Information Sciences,
						 Interventional Neuroradiology Unit, University Hospitals
						 of Geneva, Geneva 1211, Switzerland}
\affiliation{Laboratory for Hydraulic Machines,
						 Ecole Polytechnique F\'ed\'erale de Lausanne,\'e
						 1015 Lausanne, Switzerland}

\author{S. M\"uhlbauer}
\affiliation{Neutron Scattering and Magnetism, Laboratory
 							for Solid State Physics, ETH Z\"urich,
							Switzerland}

 \author{T. Giamarchi}
 \affiliation{DPMC-MaNEP, {\color{black} University of Geneva, Geneva} Switzerland}

\author{C. Kollath}
\affiliation{HISKP, Universit\"at Bonn, Nussallee 14-16,
			       D-53115 Bonn, Germany.}

 \author{A. Zheludev}
 \email{zhelud@ethz.ch}
 \homepage{http://www.neutron.ethz.ch/}
 \affiliation{Neutron Scattering and Magnetism,
              Laboratory for Solid State Physics,
              ETH Z\"urich,
							Switzerland}

\author{G. Ehlers}
\affiliation{{\color{black}Quantum Condensed Matter Division},
              Oak Ridge National Laboratory,
              Oak Ridge, Tennessee 37831, USA}

\author{A. M. Tsvelik}
\affiliation{Department of Condensed Matter Physics
 				     and Materials Science, Brookhaven National Laboratory,
 					   Upton, New York 11973-5000, USA}


\begin{abstract}
The zero-field excitation spectrum of the strong-leg spin ladder
\DIMPY~is studied with a neutron time-of-flight technique. 
The spectrum is decomposed into its symmetric and asymmetric parts with respect to the rung momentum and compared with theoretical results obtained by the density matrix renormalization group method. 
Additionally, the calculated dynamical correlations are shown for a wide range of rung and leg coupling ratios in order to point out the evolution of arising excitations, as e.g. of the two-{\color{black}magnon} bound state from the strong to the  weak coupling limit. 
\end{abstract}



\maketitle

\section{Introduction}

The discovery of novel materials as clean
realizations of quasi-one dimensional spin Hamiltonians enabled
the study of one dimensional many-body physics\cite{giamarchibook1,tsvelikbook1}
and fascinating phenomena such as
Luttinger-liquid behavior\cite{Dender1996p1,Stone2003p1,lake2005,Klanjsek2008p1}
or (quantum) phase transitions of
gapped quantum magnets\cite{Thielemann2009p1,Thielemann2009p2,Zapf2006p1,Zheludev2007p1},
in quantitative agreement with theoretical and numerical
predictions. Among these systems, the Heisenberg AF two-leg
spin-ladder \cite{Barnes1993p1} belongs to the simplest models, yet featuring non-trivial physics.
{\color{black} Recently the possibility to control such systems with the application of a magnetic field 
large enough to induce sizeable changes in the magnetization has allowed to explore 
a huge variety of novel physical phenomena\cite{Giamarchi2008p1}}.

Lately, a lot of effort was put in the study
of dimerized {\itshape strong-rung} spin-ladders,
such as e.g. the organometallic compounds \IPA ~(IPA-CuCl$_3$)
\cite{Masuda2006p1,Zheludev2007p1,Fischer2011p1} or
\Hpip ~(BPCP)\cite{Klanjsek2008p1,Ruegg2008p2,Thielemann2009p1,Bouillot2011p1}.
In this coupling limit, the zero-field
excitation spectrum is dominated by gapped but
hardly mobile dimer triplet excitations on the rung.
Nowadays, the basic underlying physics at zero magnetic field can be regarded as
well established: Analytical solutions are provided
by {\itshape e.g.} the strong-coupling
approach\cite{Sachdev1990p1,Gopalan1994p1}, starting
from non-interacting dimers.

Contrar{\color{black}il}y, the physics of {\itshape strong-leg} spin
ladders remained much more elusive, mainly
due to the lack of suitable analytic approaches in particular for the regime $J_\mathrm{rung}
/J_\mathrm{leg} \approx 1$. The existence of the
spin liquid ground state and the widely dispersive
gapped magnon is less obvious and originates in a
subtle Haldane mechanism\cite{Haldane1983p1, Dagotto1996p1}.

In contrast to the {\itshape strong-rung} limit, two-magnon
excitations become progressively more important. The strong but short-ranged
attractive potential between magnons leads to pronounced
two-magnon bound states below a two-magnon continuum. So
far, two-magnon excitations in spin-ladders were 
observed in the cuprate material
Sr$_{14}$Cu$_{24}$O$_{41}$\cite{Notbohm2007p1} and more recently,
in the organometallic low-energy scale material
\DIMPY ~(DIMPY\cite{Shapiro2007p1,Somoza2009p1}).
{\color{black} In this work, we study one- and two-magnon
excitations in the latter material with a complementary
technique and thereby extend the measurements
of Ref. \onlinecite{Schmidiger2011p1} and \onlinecite{Schmidiger2012p1}}.


DIMPY is currently the cleanest\footnote{The inter-ladder
interaction was found to be $J'\approx 6\,\mu\mathrm{eV}$,
more than 2 orders of magnitude smaller than $J_\mathrm{leg}$
and $J_\mathrm{rung}$.} realization of a {\itshape strong-leg}
spin-ladder material. It crystallizes in a monoclinic structure
with space group P$2(1)/$n and lattice constants\cite{Shapiro2007p1}
$a = 7.504$\AA~, $b = 31.61$\AA~, $c = 8.202$\AA~ and
$\beta = 98.98^\circ$. A depiction of the unit cell content can be
found in Ref. \onlinecite{Schmidiger2011p1}. Cu$^{2+}$ ions with
an effective spin $S=1/2$ in a tetrahedral environment of Br$^{-}$
ions are interacting through Cu-Br-Br-Cu superexchange pathways,
thereby building a ladder-like spin-network. Dimpy features two
different ladder systems, both running along the crystallographic
${\bf a}$-axis but being described by distinct rung vectors
${\bf d}_{1,2}=(0.423,\pm0.256,0.293)$, in fractional coordinates
\cite{Shapiro2007p1}. Recent zero-field triple-axis neutron
scattering experiments in combination with DMRG calculations
{\color{black} indicated that the low-energy physics is governed by the 
Heisenberg spin-ladder Hamiltonian}

\begin{align}
\mathcal{H} = J_\mathrm{leg} \sum_{l}\sum_{j=1}^{2}{\bf S}_{l,j}\!\cdot\!{\bf S}_{l+1,j} + J_\mathrm{rung} \sum_{l} {\bf S}_{l,1}\!\cdot\!{\bf S}_{l,2}.
\end{align}
{\color{black} Neutron experiments in combination with PCUT calculations\cite{} estimated 
$J_\mathrm{leg}/J_\mathrm{rung}\approx 2.2(2)$,  while careful measurements of the magnon 
dispersion \cite{Schmidiger2011p1} 
in combination with density matrix renormalization group 
(DMRG) calculations\cite{Schmidiger2012p1} determined the exchange constants to be $J_\mathrm{leg}\!=\!1.42(6)$~meV
and $J_\mathrm{rung}\!=\!0.82(2)$~meV. Additional intra-ladder interactions were found to
be insignificant\cite{Schmidiger2011p1} while low-temperature specific 
heat measurements\cite{Schmidiger2012p1} estimated inter-ladder interactions 
to be on the order of $6\,\mu\mathrm{eV}$.}
In the following, the theoretical calculations are performed using the time-dependent DMRG method with $J_\mathrm{rung}$ and $J_\mathrm{leg}$ as
quoted above. For details on the calculations we refer to Ref.~\onlinecite{Schmidiger2012p1}.

Due to the absence of e.g. diagonal interactions, the spin Hamiltonian
possesses leg-permutation symmetry and the total dynamical structure factor {\color{black}$\mathcal{S}(\mathbf{q},\omega)$}
decomposes into a symmetric {\color{black} $\mathcal{S}_{+}(\mathbf{q},\omega)$}
and asymmetric {\color{black}$\mathcal{S}_{-}(\mathbf{q},\omega)$}
part\cite{Schmidiger2011p1},
 \begin{eqnarray}
 \mathcal{S}(\mathbf{q},\omega) & = & s^{+}(\mathbf{q})\,\mathcal{S}_{+}(\mathbf{q},\omega)+ s^{-}(\mathbf{q})\, \mathcal{S}_{-}(\mathbf{q},\omega)\label{decom},
 \end{eqnarray}
where $s^{-}(\mathbf{q})$ and $s^{+}(\mathbf{q})$
denote the asymmetric and symmetric structure factor respectively. Assuming the two ladder systems with $\mathbf{d}_{1,2}$
to be non-interacting, {\color{black} they} are given by

\begin{align}
4\,s^{\pm}({\bf q}) = 2 \pm \cos({\bf q}\cdot {\bf d}_1) \pm \cos({\bf q}\cdot {\bf d}_2).
\label{sf}
\end{align}

Odd and even number of magnon excitations contribute
to the  asymmetric and symmetric channel respectively.

In neutron scattering experiments, the partial differential
cross section is measured. For magnetic scattering as
discussed in this work, it is given by \cite{squiresbook}

\begin{align}
\frac{\mathrm{d}^2\sigma}{\mathrm{d}\Omega \mathrm{d}\omega} \propto N \frac{k_f}{k_i} |F({\bf q})|^2 \mathcal{S}(\mathbf{q},\omega),
\end{align}
where $N$ denotes the number
of unit cells in the sample, {\color{black} $F(\mathbf{q})$ the magnetic form factor, ${\bf k}_i$ (${\bf k}_f$) 
the wavevector of the incident (final) neutrons and $\mathbf{q} = \mathbf{k}_f-\mathbf{k}_i$ the 
momentum transfer. The latter can be written as
$\mathbf{q} = h\mathbf{a}^\star + k\mathbf{b}^\star+l\mathbf{c}^\star$
with $\mathbf{a}^\star, \mathbf{b}^\star$ and $\mathbf{c}^\star$ describing the 
reciprocal lattice vectors of the crystal.} Due to the
different structure factors of the symmetric and asymmetric
channel, {\color{black} symmetric and asymmetric excitations} can be {\color{black} fully} separated in a neutron scattering experiment.

In recent experiments, the single magnon dispersion
was measured by triple-axis neutron scattering and found
to be persisting throughout the complete Brillouin zone,
conf{\color{black}i}rming the leg-permutation symmetry \cite{Schmidiger2011p1}.
In subsequent triple-axis experiments, a two-magnon bound state
was observed by performing scans at three positions in reciprocal
space,  $(h,k,l)=(\eta,0,-1.44 \cdot \eta)$ with $\eta=0.5,0.625$ and
$0.75$, quantitatively confirming numerical density matrix
renormalization group (DMRG) calculation of $\mathcal{S}_{+}
(\mathbf{q},\omega)$\cite{Schmidiger2012p1}.

{\color{black} The goal of this work is twofold: First, we extend the 
measurements of Ref. \onlinecite{Schmidiger2011p1} and
\onlinecite{Schmidiger2012p1} by using the complementary neutron time-of-flight technique, since a detailed analysis of the symmetric and asymmetric 
zero-field spectrum of DIMPY has not yet been performed.
We study the bound state in detail, definitely
prove the leg-permutation symmetry and separate the symmetry 
channels completely. Secondly, DMRG calculations of the dynamical 
structure factor were performed for different coupling ratios $0.5 < J_\mathrm{leg}/J_\mathrm{rung} < \infty$. This enables us to 
numerically observe the evolution of excitations from the 
strong-rung to the strong-leg regime and to compare it to
existing analytic results.}

\section{Experiment}

For the present experiment, the same sample
as in Ref.~\onlinecite{Schmidiger2011p1},\onlinecite{Schmidiger2012p1}
was used. It consisted of four fully deuterated single crystals \DIMPYd~ with a
total mass of 3.7~g and co-aligned to a mosaic spread better than
1.5$^\circ$. Measurements were performed at the CNCS cold neutron
chopper spectrometer \cite{Ehlers2011p1} at SNS spallation source.
Temperature was controlled with a conventional $^4$He-cryostat and
the sample was mounted with the ${\bf b}$-axis vertical. Measurements
were performed at $T=1.5$~K and background data was collected at 50~K
and 110~K. The incident energy was fixed to $4.2$~meV and
the sample was rotated by 180$^\circ$ in steps of 5$^\circ$.
Intensity was normalized to the proton charge on the target:
$1.5\mu \mathrm{C}$ {\color{black} (40 minutes counting time)} per {\color{black} rotation} step for
the 1.5~K and $0.75\mu \mathrm{C}$ for the 50~K and
110~K measurement respectively.

\section{Results and Discussion}

\subsection{Integrated intensity and structure factor}

\begin{figure}[h!t]
\includegraphics[width=0.95\columnwidth]{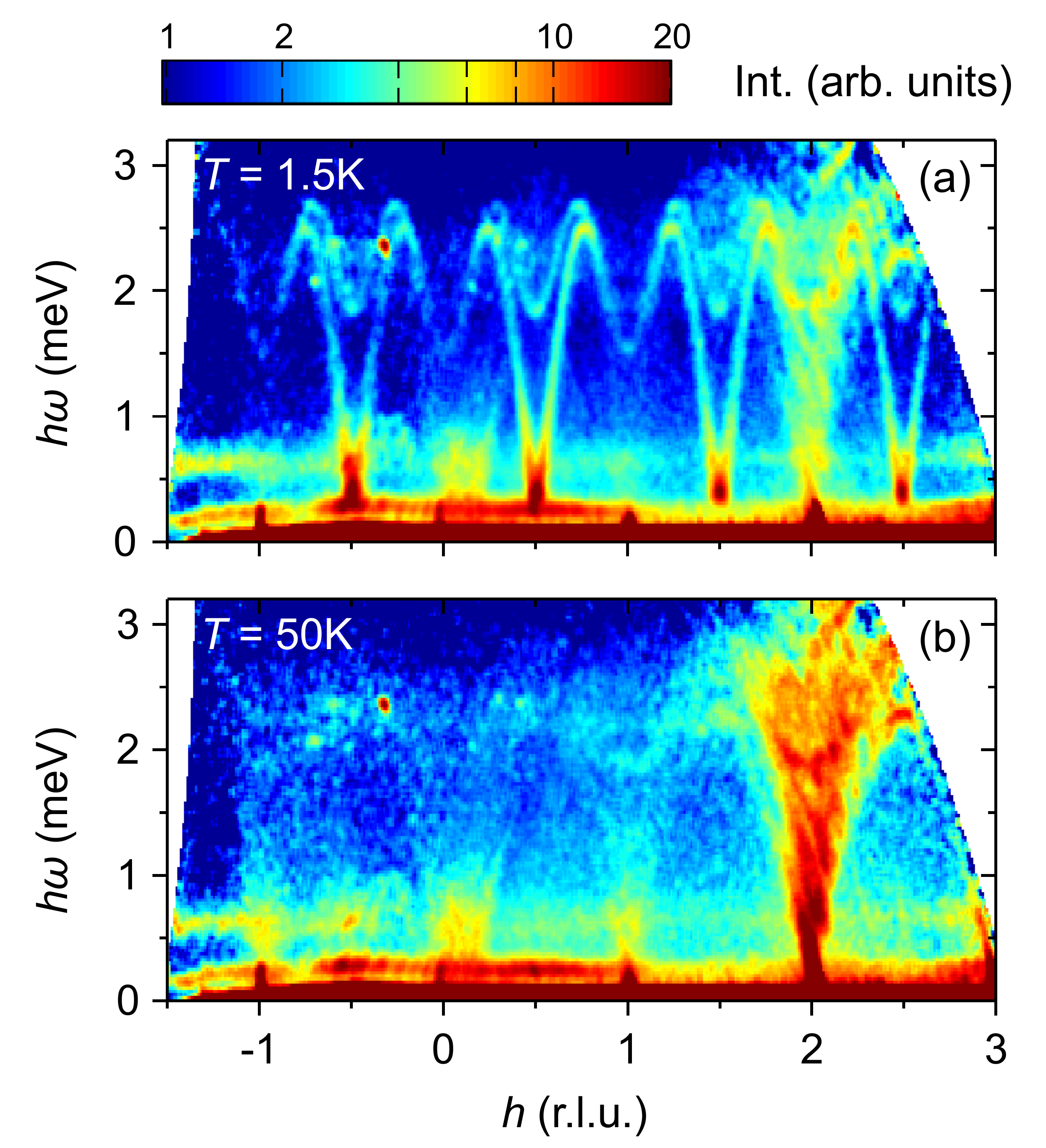}
\caption{(Color online) Raw data from neutron time-of-flight experiments in
												(C$_7$D$_{10}$N)$_2$CuBr$_4$, measured at $(a)$ 
												$T=1.5$~K and $(b)$ $T=50$~K. Data was integrated along the 
												non-dispersive ${\bf b}^\star$ and ${\bf c}^\star$ direction.
												{\color{black}Intensity is shown as a function of energy 
										    transfer $\hbar\omega$ and momentum transfer along the leg $h$,
										    in reciprocal lattice units.}} \label{raw}
\end{figure}

Due to the exceptional one-dimensional nature of DIMPY,
no dispersion along the perpendicular directions
${\bf b}^\star$ and ${\bf c}^\star$ was observed
previously\cite{Hong2010p2}. {\color{black} Neutron time-of-flight} data can hence be
integrated along these directions, thereby
improving statistics. Raw data at $T=1.5$~K
and 50~K, integrated along ${\bf b}^\star$
and ${\bf c}^\star$ using the Horace
program\cite{Horace} is shown in figure
 \ref{raw}a,b.

At $1.5$~K both the single-magnon excitation
and the two-magnon bound state are observed
over four Brillouin zones. However, the
magnetic signal is contaminated by $T$-dependent
and $T$-independent contributions. The $T$-dependent contributions are mainly due to the inelastic phonon scattering. {\color{black} This is in contrast to} $T$ independent background, which can stem from both coherent and incoherent scattering by the sample and equipment. Such contributions are evident in comparison with the measurement at 50~K (fig. \ref{raw}b).

Background subtraction was performed taking both of these contributions into ac{\color{black}c}ount as described
in Appendix A, using the integrated data sets.
In figure \ref{sub}, background subtracted data
is shown in the Brillouin zone $0\!<\!h\!<\!1$.
Clearly, most of the background features are
removed by our procedure.


\begin{figure}[h!t]
\includegraphics[width=1\columnwidth]{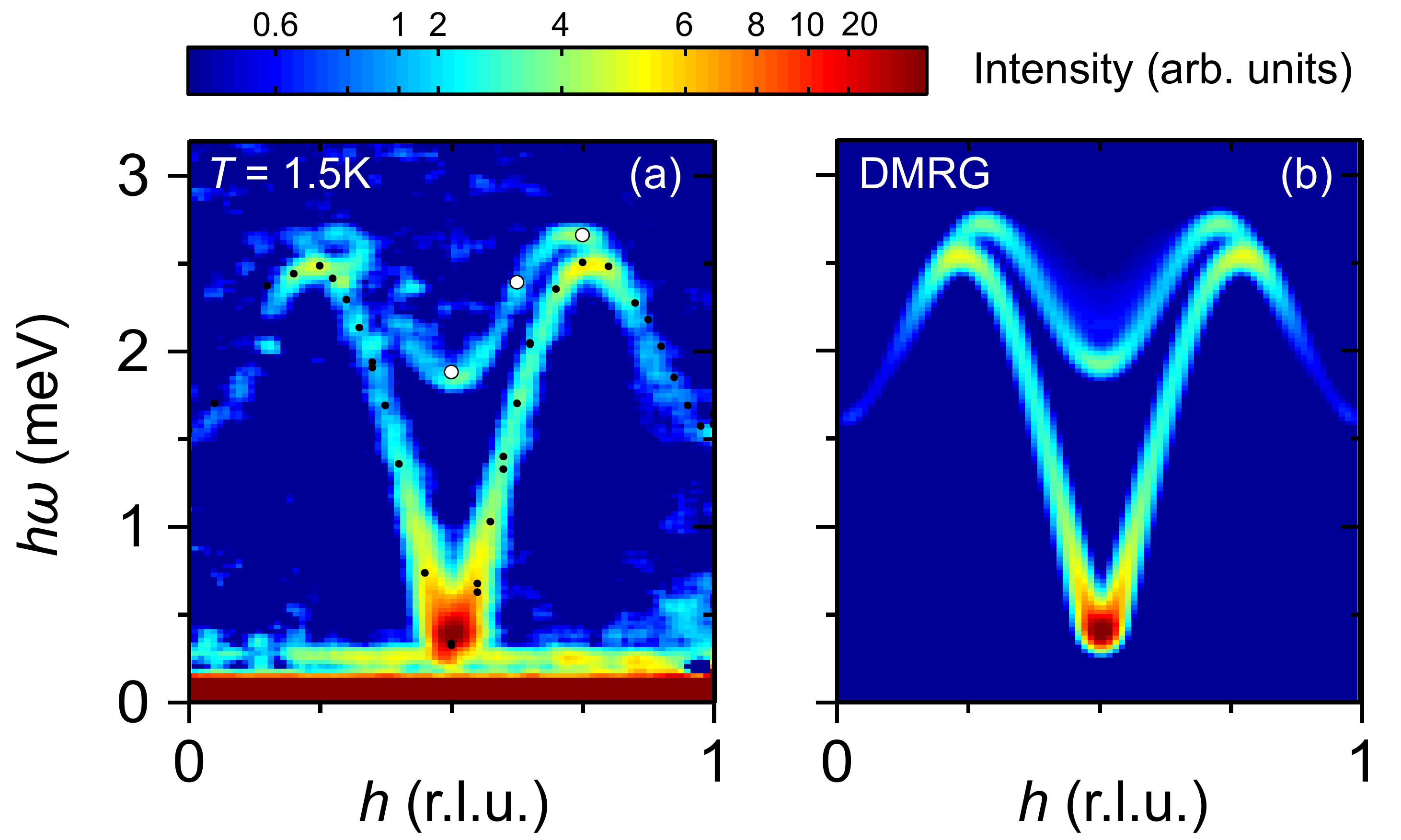}
\caption{(Color online) $(a)$ Background subtracted data from neutron time-of-flight experiments
												in (C$_7$D$_{10}$N)$_2$CuBr$_4$, integrated along
												the ${\bf b}^\star$ and ${\bf c}^\star$ direction.
												{\color{black}Intensity is shown as a 
												function of energy transfer $\hbar\omega$ and 
												momentum transfer along the leg.} Black and white points correspond to triple-axis
												measurements of the magnon and two-magnon bound state excitations (Ref.
												 \onlinecite{Schmidiger2011p1,Schmidiger2012p1}). $(b)$ Numerical DMRG calculation of the
												dynamical structure factor $\mathcal{S}_{+}(h,\omega)
												+ \mathcal{S}_{-}(h,\omega)$, convoluted with experimental
												resolution. {\color{black} The measured and calculated
												spectra show remarkable agreement.}} \label{sub}
\end{figure}


Due to the integration process, the cosine functions
in equation (\ref{sf}) average to zero and an equal
combination of $\mathcal{S}_{+}(h,\omega)$ and
$\mathcal{S}_{-}(h,\omega)$ is observed. The measured
magnon dispersion agrees with the recently performed
triple-axis experiment at $T=50$~mK (black points,
from Ref. \onlinecite{Schmidiger2011p1}). {\color{black} Moreover},
the two-magnon
bound state clearly persists in the region $0.2<h<0.8$
and is consistent with the three constant-${\bf q}$
scans performed in Ref. \onlinecite{Schmidiger2012p1}
(white points), while two-magnon continuum excitations
are too weak to be observed under experimental conditions.
The numerical calculation of $\mathcal{S}_{+}(h,\omega)+
\mathcal{S}_{-}(h,\omega)$ convoluted with an approximate
experimental resolution (Fig. \ref{sub}b) is in quantitative
agreement with the experiment, both in terms of dispersion
and intensity of the two sharp modes.

\begin{figure}[tbp]
\includegraphics[width=1\columnwidth]{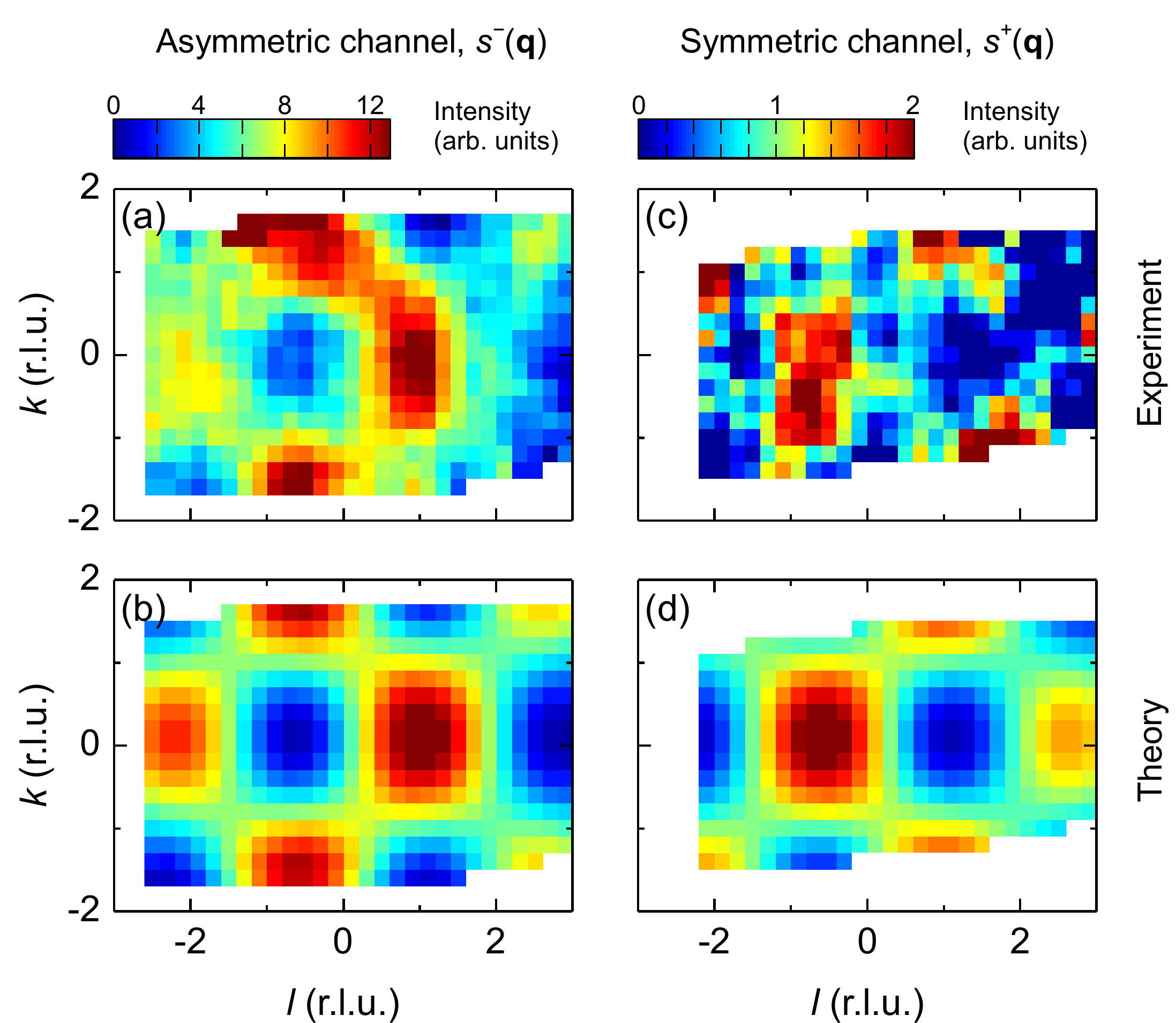}
\caption{(Color online) Background subtracted data $\mathcal I^{\mathrm{sub}}_{h,\omega}(k,l)$,
					as a function of momentum transfer along the perpendicular directions ${\bf b}^*$ and
					${\bf c}^*$, in order to visualize $(a)$ the asymmetric and $(b)$ the symmetric structure
					factors $s^{\pm}_{h}(k,l)$. Data was integrated along $h=[0.45,0.55]$~rlu and in the energy range
					$(a)$ $\hbar\omega=[0.2,0.6]$~meV  and $(c)$ $[1.75,2.05]$~meV.
					The corresponding calculation of the asymmetric and symmetric structure factor
					$s^{\pm}_{h}(k,l)$ weighted by the magnetic form factor are shown in
					$(b)$ and $(d)$.} \label{struc}
\end{figure}

In order to map out the structure factors
$s^{\pm}({\bf q})$, raw data at 1.5~K, 50~K
and 110~K was integrated around the magnetic
zone center $h = [0.45,0.55]~$~rlu and in
the energy range $\hbar\omega=[0.2,0.6]$~meV
and $[1.75,2.05]$~meV, enclosing the magnon
and two-magnon bound states, respectively. As described
in Appendix A, the instrumental and phonon
background were separated, extrapolated to
1.5~K and subtracted (fig. \ref{struc}a,c).


The structure factors $s^{\pm}_{h}(k,l)$ given
by equation (\ref{sf}) were calculated for $h=0.5$
on the same $k,l$-grid as experimental data (fig.
\ref{struc} b,d). They were multiplied by the magnetic
form factor of the Cu$^{2+}$ ion, given by
$|F_h(k,l)|^2\approx |\langle j_0 \rangle_h(k,l)|^2$
and with the function $j_0$ as numerically calculated
in Ref. \onlinecite{magneticformfactor1}.

The cut around the single magnon excitation (fig.
\ref{struc}a) clearly follows the predicted asymmetric
channel structure factor  $s^{-}_{h}(k,l)$ (fig.
\ref{struc}b). Allthough the vertical coverage of
a 2D time-of-flight detector is limited by
$\pm 16^\circ$, the exceptionally long $\bf b$-axis of
DIMPY ($b\!=\!31.61$~\AA) enables to observe a full
period of the structure factor in vertical direction.

{\color{black} Although the signal for the cut around the two-magnon bound state
(fig. \ref{struc}c) is much weaker, we observe enhanced intensity
in the range $-2\!<l\!<0$ with $-1\!<k\!<1$ while it is
basically zero for $0\!<l\!<2$. This is in agreement with
the calculated variation of the symmetric structure factor
$s^{+}_{h}(k,l)$ as shown in fig. \ref{struc}d. The assumption 
of two non-interacting ladder systems with structure factors
as in equation (\ref{sf}) is hence found to be valid.}

\subsection{Channel Separation}

As a next step, the two symmetry channels were separated.
The basic idea was to divide the 4D data set of 2D cuts
by integrating along small ranges in $h$ and
$\hbar \omega$ and to determine the contribution of the
asymmetric and symmetric channel for each value of $h_i$
and $\hbar \omega_i$, assuming that the structure factors
are given by equation (\ref{sf}).

The $h$-$\hbar\omega$-plane was divided into 1600 boxes
$(h_i,\hbar\omega_i)$ of the size
$0.025\,\mathrm{rlu}\times 0.075\,\mathrm{meV}$. For
each box $(h_i,\hbar\omega_i)$, data measured at
$1.5$K, $50$K and $110$K was integrated along a small
range $h_i\pm0.025$~rlu and $\hbar\omega_i\pm 0.075$~meV,
leaving 2D data sets $\mathcal I^{T}_{h_i,\omega_i}(k,l)$.
Background subtraction was performed for each
$(h_i,\hbar\omega_i)$ using the data sets
$\mathcal I^{T}_{h_i,\omega_i}(k,l)$ with $T=1.5$~K,
50~K and 110~K, as described in Appendix A.

The structure factor for the
asymmetric and symmetric channel $s^{\pm}_{h_i}(k,l)$
were calculated on the same grid as the data.
Two masks $M^\pm_{h_i,\omega_i}(k,l)$ were defined by

\begin{align}
M^\pm_{h_i,\omega_i}(k,l)=\begin{cases}
  s^{\pm}_{h_i,\omega_i}(k,l)^{-1} |F_{h_i}(k,l)|^{-2}  & s^\pm_{h_i,\omega_i}(k,l)\ge L_\pm \\
  0 & \mathrm{else}
\end{cases}_,
\end{align}

such that the asymmetric and symmetric masks
$M^\pm_{h_i,\omega_i}(k,l)$ cut out data in
the region where intensity from the corresponding
channel is expected by the structure factor. The
threshold for the channel were taken to be $L_+ = 0.8$
and $L_- = 0.85$, respectively\footnote{The threshold
for the symmetric channel was taken to be slightly lower
in order to resolve the two-magnon bound state which is much weaker than
the magnon excitations.}.

In order to determine the asymmetric contribution,
background subtracted data
$\mathcal I^{\mathrm{sub}}_{h_i,\omega_i}(k,l)$ was
multiplied elementwise by the asymmetric mask
$M^-_{h_i,\omega_i}(k,l)$, summed up and divided by the number
of non-zero elements - leaving one number $I^{-}(h_i,\omega_i)$
describing the asymmetric contribution at the position
$(h_i,\omega_i)$. The same procedure was performed for the symmetric
mask $M^+_{h_i,\omega_i}(k,l)$ leading to the symmetric contribution
$I^{+}(h_i,\omega_i)$ at $(h_i,\omega_i)$.


\begin{figure}[tbp]
\includegraphics[width=1\columnwidth]{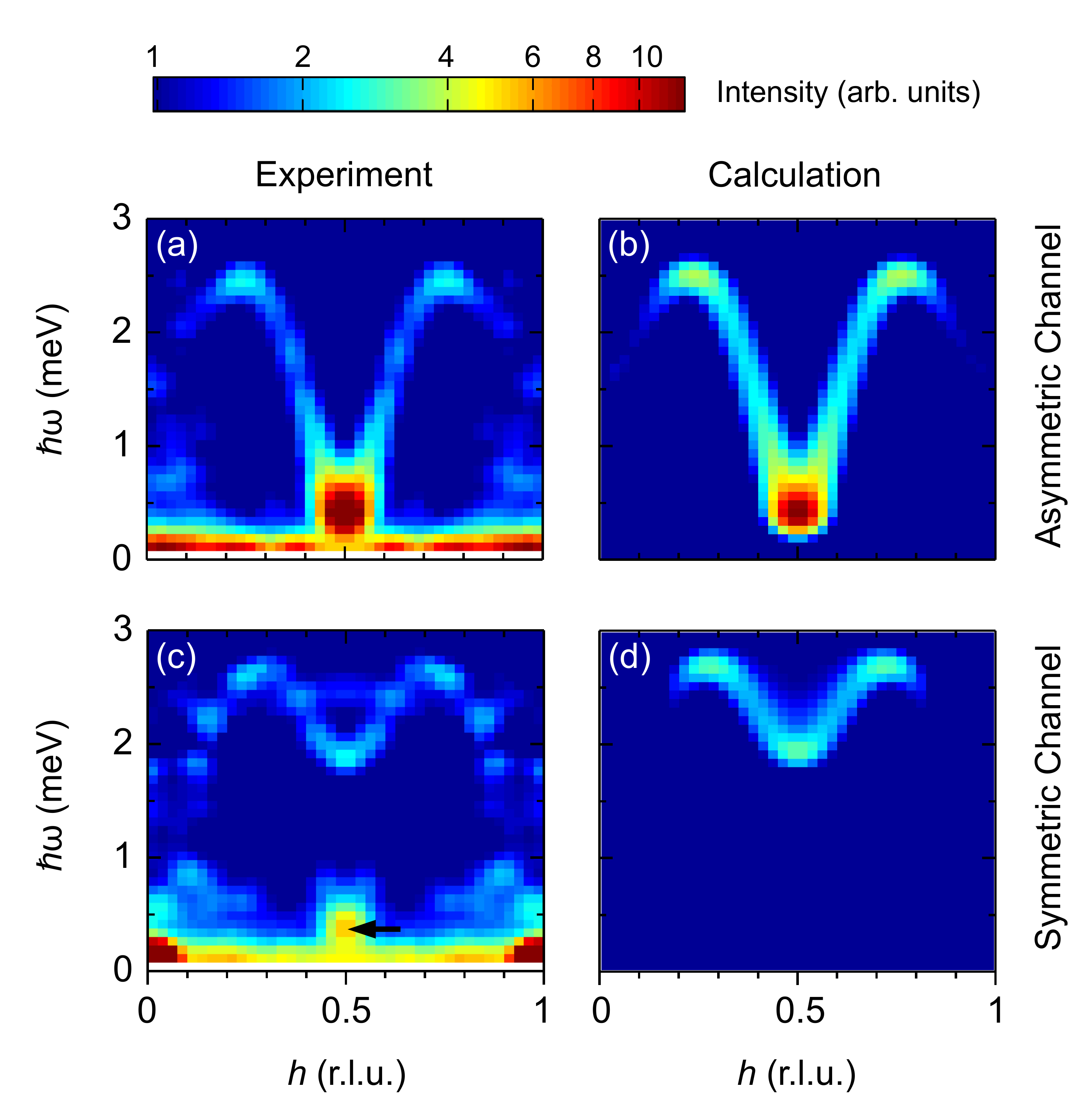}
\caption{(Color online)  Separation of the asymmetric and
					symmetric channel in \DIMPYd. $(a)$
					asymmetric and $(c)$ symmetric contribution extracted 				
					from raw data as explained in the text. A spurious 
					remainder of the single-magnon excitation
					is still visible in the separated symmetric
					contribution $(c)$ (black arrow).
					DMRG calculation of $(b)$ $\mathcal{S}_{-}(h,\omega)$
					and $(d)$ $\mathcal{S}_{+}(h,\omega)$,
					convoluted with a similar resolution as in $(a)$.
					} \label{sep}
\end{figure}


The separated and symmetrized contributions
$I^-(h,\omega)$ and $I^+(h,\omega)$ in the first
Brillouin zone $0\!<\!h\!<\!1$ are shown in figure
\ref{sep}a,c respectively. $I^-(h,\omega)$ clearly
contains the single-magnon excitation while no
contribution from the two-magnon bound state
is visible. Figure \ref{sep} shows the DMRG
calculation of $\mathcal{S}_{-}(h,\omega)$, convoluted
with a similar resolution as in fig. \ref{sep}a.
However, due to the separation process, the
intensity cannot be compared anymore. The two-magnon
bound state is clearly visible in the separated even
channel $I^+(h,\omega)$, in agreement with the
calculated $\mathcal{S}_{+}(h,\omega)$ (fig. \ref{sep}d).
$I^+(h,\omega)$ still contains a 'ghost' of the single-magnon
excitation (black arrow) an artefact of the separatation process.

\section{Discussion}

The results can be summarized as follows. 
(1) The excitations spectrum of DIMPY is dominated by the well-known magnon excitation as well as a
strong and highly dispersive two-magnon bound state,
persisting throughout about 60\% of the Brillouin zone.
The intensity and dispersion of both modes are in
full agreement with the DMRG calculations.
(2) The symmetric and asymmetric structure factor follows
the prediction for non-interacting ladder systems with
rung vectors $\mathrm{d}_{1,2}$ and can directly be
mapped out in a TOF experiment.
(3) DIMPY features the leg-permutation symmetry. The
asymmetric and symmetric excitation channel can
be fully separated. The former contains the magnon
excitation while the latter contains the two-magnon
bound state.

In order to put results into context, we show in the following DMRG
calculations of the momentum- and frequency resolved dynamical
structure factor in both symmetry channels from the strong-rung to the strong-leg regime.
This enables us to relate to numerous aspects of the spin-ladder problem in either
coupling regimes which were studied in detail before, both analytically
and numerically{\color{black}\cite{Dagotto1996p1,Damle1998p1,Greven1996p1,
Sushkov1998p1,Zheng2001p1,Reigrotzki1994p1,Oitmaa1996p1,Knetter2001p1}}. 

\begin{figure}[tbp]
\includegraphics[width=1\columnwidth]{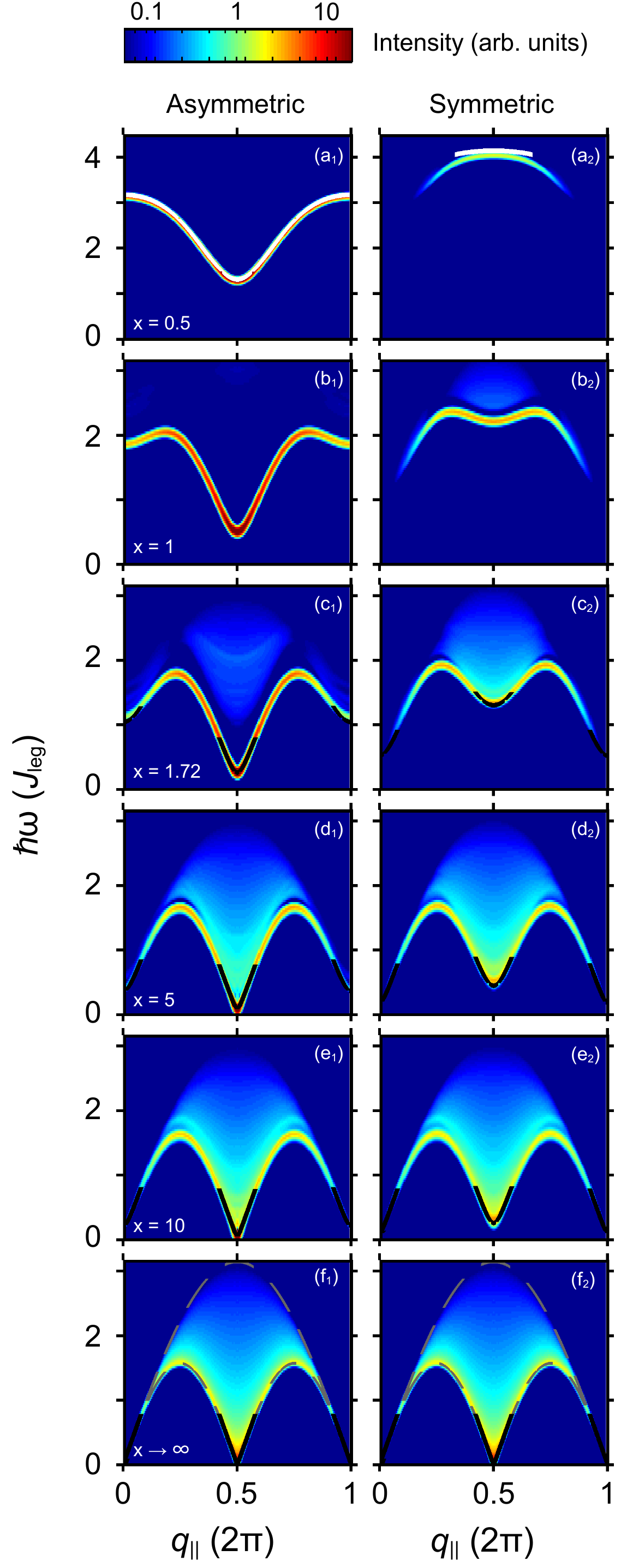}
\caption{(Color online) DMRG calculation of the momentum- and
frequency resolved dynamic structure factor in the asymmetric
(left) and symmetric (right) channel with rung momentum
$q_\perp = \pi$ and $0$, respectively. White lines correspond to
predictions from Ref. \onlinecite{Zheng2001p1,Reigrotzki1994p1}. Black lines indicate
the boundaries of multi-particle continua as described in
the text.}
\label{evolution}
\end{figure}

The calculations were performed for different coupling ratios $x =
J_\mathrm{leg}/J_\mathrm{rung}$, particularly for  $x = 0.5, 1,
1.72, 5, 10$, with $J_\mathrm{leg}$ fixed to unity, as well as
for the spin chain ($x\rightarrow \infty$). Figure \ref{evolution}
shows the calculated structure factor in the asymmetric 
({\itshape left}) and symmetric ({\itshape right}) channel for 
different coupling ratios. {\color{black} It is shown as a function of energy
$\hbar\omega$ and momentum along the leg of the ladder, 
$q_\parallel := \bf{q\cdot a}$.}

In the strong-rung regime with {\color{black} $x < 1$} (fig. \ref{evolution}a,b),
the spin ladder consists of weakly interacting dimers. The asymmetric
dynamical structure factor is dominated by a hardly dispersive magnon
excitation with a gap $\Delta \simeq J_\mathrm{rung} - J_\mathrm{leg}$
and a bandwidth $W \simeq 2 J_\mathrm{leg}$. The symmetric channel
contains a weak two-magnon continuum as well as a $S=1$ bound state existing
in a narrow region around the magnetic zone center $q_\parallel = \pi$.

For strong-rung ladders, the dispersion \cite{Reigrotzki1994p1} {\color{black} and}
the spectral weight of the magnon, the two-magnon continuum
as well as the bound states were calculated using the strong-coupling
expansion\cite{Sushkov1998p1,Damle1998p1,Oitmaa1996p1} and
with the linked cluster series expansion\cite{Zheng2001p1}.
White lines in figure \ref{evolution}a correspond to the
calculation in Ref. \onlinecite{Reigrotzki1994p1,Zheng2001p1} and
agree for $x = 0.5$. Moreover, the results indicate that for
{\color{black} $x  \rightarrow 0$}, the triplet
bound state exists only in a narrow q-range with
{\color{black} $2\pi/3\!<\!q_{\parallel}\!<\!4\pi/3$}
(i.e. in $1/3$ of the Brillouin zone) and that
the spectral weight of the triplet state scales with $x^2$ (Ref.
\onlinecite{Damle1998p1}). In strong-rung ladders, two-magnon
excitations are hence usually too weak to be observable by neutron
scattering methods.

In contrast, for the strong-leg coupling regime {\color{black} $x>1$} both the 
symmetric and asymmetric dynamical structure factors converge towards
the two-spinon continuum excitation spectrum
of the Heisenberg $S=1/2$ spin chain {\color{black} for $x \rightarrow \infty$ (fig. \ref{evolution}f).
The latter is gapless, features a bandwidth of $W=\pi J_\mathrm{leg}$ and}
is bounded\cite{Muller1981p1} by $\epsilon_\mathrm{l} = \pi J_\mathrm{leg}/2 |\sin{(q_\parallel)}|$
and $\epsilon_\mathrm{u} = \pi J_\mathrm{leg} |\sin{( q_\parallel/2)}|$ {\color{black}(grey broken
lines in fig. \ref{evolution}f)}.

{\color{black} At finite inter-chain interaction $J_\mathrm{rung}$, 
spinons are confined and asymmetric excitations aquire a spin gap of
$\Delta \simeq 0.41 J_\mathrm{rung}$ if $x \gg 1$
(Ref. \onlinecite{Greven1996p1})}. As pointed out
by Shelton et al. \cite{Shelton1996p1}, the bosonized Hamiltonian can
be mapped onto a system of weakly interacting massive
triplet and singlet Majorana fermions, with masses
$m_s \simeq 3 m_t$ and with the velocity of the Heisenberg spin chain,
$v = \pi J_\mathrm{leg} /2$. The triplet Majorana fermion is asymmetric
under leg permutation symmetry (its rung momentum is $q_\perp = \pi$)
while the singlet Majorana fermion is symmetric with $q_\perp = 0$.
The dynamic structure factor contains various sharp and continous single and
multi Majorana fermion excitations, summarized in table \ref{table:exc}.


\begin{table}[h!t]
\begin{center}
\begin{tabular}{p{2cm}p{1cm}p{1cm}p{2cm}}
\hline \hline
Excitation & $q_\perp$ & $q_{\parallel,\mathrm{min}}$ & Threshold $m_\mathrm{tresh}$\\
\hline
1T 				&  $\pi$ 	&	$\pi$ &	1$m$ \\
2T		    &	 $0$	  & $0$ 	& 2$m$ \\
3T			  &	 $\pi$  & $\pi$ & 3$m$ \\
1T + 1S   &  $\pi$  & $0$   & 4$m$ \\
2T + 1S   &  $0$    & $\pi$ & 5$m$ \\
\hline
\end{tabular}
\caption{Multi triplet (T) and singlet (S) low energy excitations. The rung
$q_\perp$ denotes the symmetry channel in which these excitations appear while
$q_{\parallel,\mathrm{min}}$ determines whether they occur around $\pi$ or $0$. $m_\mathrm{thres}$
describes to the gap of the corresponding excitation.}
\label{table:exc}
\end{center}
\end{table}


The triplet excitation (the ``magnon'') is a sharp mode around $q_\parallel = \pi$
only and its dispersion is described by
\begin{align}
\epsilon_t = \sqrt{m^2 + v^2 (q_\parallel - \pi)^2}
\end{align}
{\color{black} with $m\approx 0.41\,J_\mathrm{rung}$}. The lower boundaries of the multi-particle continua are given by
\begin{align}
\epsilon_\mathrm{l} = \sqrt{m_\mathrm{thres}^2 + v^2 (q_\parallel - q_{\parallel,\mathrm{min}})^2}
\end{align}
and can be observed either in the symmetric ($q_\perp = 0$) or asymmetric ($q_\perp = \pi$)
channel. These predicitions are shown as {\color{black} black} full lines in figure
\ref{evolution}c-f and successfully describe the lower boundary of
the dynamical structure factor in the strong-leg regime, confirming
the analytical predictions. {\color{black}In particular in the symmetric channel, 
the continuum around $q_\parallel=\pi$ corresponds to two triplet 
and one singlet excitation, whereas the continuum around 
$q_\parallel=0$ results from two triplet excitations.}

{\color{black} For the experimental value $x \approx 1.72$ {\color{black} in DIMPY} instead of the threshold singularity the symmetric channel displays a rather well defined coherent mode. In the framework of Ref. \onlinecite{Shelton1996p1} this corresponds to the three-particle bound state of two triplet and one singlet excitation. This value of $x$ is too small for the above theory to yield good quantitative predictions, though it contains a provision for such bound state. This provision comes in the form of the interaction the Majorana fermions originating from the coupling of uniform magnetizations of the two chains\cite{Shelton1996p1}. The massive Majorana fermions interact and with the proper sign of interaction can create bound states.} {\color{black} Nevertheless, in contrast to strong-rung spin
ladders, this bound state is only 8 times weaker than the single magnon excitation at $q_\parallel=\pi$. Moreover, the two-magnon bound state seems not confined to a narrow q-range but persists throughout about 60\% of the Brillouin zone and shows itself a structured dispersion. The latter does not yet follow the expansion based on the strong-coupling approach in Ref. \onlinecite{Zheng2001p1}.}

{\color{black} Being in an intermediate coupling limit with neither $x\ll1$ nor $x\gg1$, 
the observed bound state around $q_\parallel = \pi$ in DIMPY can hence be  understood {\itshape qualitatively} either as a bound state of two dimer-triplet excitations or a bound state of two-triplet and one singlet Majorana Fermion excitations in the language of Ref. \onlinecite{Shelton1996p1}, although analytic solutions from both coupling limits can not describe the bound-state {\itshape quantitatively} anymore. 
}

\section{Conclusion}

In conclusion, detailed follow-up zero-field
measurements on the strong-leg spin ladder
material DIMPY were performed by the neutron
time-of-flight technique. The two-magnon bound
state recently observed by the triple-axis
scattering technique was studied in detail and
shown to be persisting throughout 60\% of the
Brillouin zone. The structure factor of the
even and odd excitation channel was measured
and shown to be consistent with the model of
two non-interacting ladder systems described
by the rung vectors ${\bf d}_{1,2}$. It was
shown how the large 4D data set collected in
a time-of-flight experiment can be used in a
smart way in order to separate the two channels.
Moreover, the evolution of the dynamical structure
factor in the strong-leg regime was studied and
it was shown how both the symmetric and
asymmetric channel converge towards the two-spinon
continuum of a spin-chain.


\begin{acknowledgements}
This work is partially supported by the
Swiss National Fund under division II and
through Project 6 of MANEP. {\color{black} Research at Oak
Ridge National Laboratory's Spallation Neutron
Source was supported by the Scientific User
Facilities Division, Office of Basic Energy
Sciences, U. S. Department of Energy. We thank
Dr. Andrey  Podlesnyak and the SNS sample environment
team for their support during the experiment. }
\end{acknowledgements}

\subsection{Appendix A: Background subtraction}

In the following Appendix, we briefly describe the
background subtraction procedure. It is a standard
approach and was performed in a similar way in e.g.
Refs. \onlinecite{Dender1996p1}, \onlinecite{Stone2006p1}.
For the present experiment, two sources of background
were assumed: $(1)$ Temperature-independent background
both from the cryostat and other equipment as well as
coherent and incoherent scattering from the sample
\cite{squiresbook} and $(2)$ inelastic phonon scattering from
the sample, proportional to the bose-factor
$n(\omega)+1$. The total signal {\color{black}$\mathcal{I}({\bf Q},\omega, T)$}
at $T_1 = 50$~K and $T_2 = 110$~K was modelled as


\begin{align}
\mathcal I({\bf Q},\omega, T) = & \mathcal A({\bf Q}, \omega) + \mathcal B({\bf Q}, \omega) (n(\omega,T)+1)
\end{align}


with $n(\omega) = (\mathrm{e}^{\hbar\omega/k_\mathrm{B}T}-1)^{-1}$
and  $\mathcal A$, $\mathcal B$ describing the T-independent and
T-dependent background, respectively. The background contributions
can be calculated by


\begin{align}
\mathcal{B}({\bf Q},\omega) &= \frac{\mathcal I_1({\bf Q},\omega)- \mathcal I_2({\bf Q},\omega)}{n(\omega,T_1)-n(\omega,T_2)}\\
\mathcal{A}({\bf Q},\omega) &= \mathcal I_1({\bf Q},\omega)- \mathcal B({\bf Q},\omega) (n(\omega,T_1) + 1).
\end{align}


The background subtracted signal at base temperature $T_0=1.5$~K is therefore
\begin{align}
\mathcal I^\mathrm{sub}({\bf Q},\omega) = \mathcal I_{\mathrm{0}}({\bf Q},\omega) &- \mathcal{A}({\bf Q},\omega) \notag\\
&- \mathcal{B}({\bf Q},\omega)  (n(\omega,T_0) + 1).
\end{align}


\end{document}